\documentclass[%
superscriptaddress,
 amsmath,amssymb,
prl,
twocolumn,
]{revtex4-2}
\usepackage[export]{adjustbox}
\usepackage{graphicx}
\usepackage{color}
\usepackage{xcolor}
\usepackage{natbib}
\usepackage{wasysym}
\usepackage{dcolumn}
\usepackage{bm}
\usepackage{hyperref}
\definecolor{mygreen}{rgb}{0,0.5,0}
\definecolor{mybrown}{rgb}{0.65,0.16,0.16}

\def\beq {\begin{equation}}
\def\eeq {\end{equation}}
\def\beqa {\begin{eqnarray}}
\def\eeqa {\end{eqnarray}}
\def \bnum {\begin{enumerate}}
\def \enum {\end{enumerate}}
\def\bi {\begin{itemize}}
\def\ei {\end{itemize}}

\def\rel {R_{\lambda}}

\def\la {\langle}
\def\ra {\rangle}

\def\mbf {\mathbf}

\def\epsm{\la{\epsilon}\ra}
\def\nr{r/\eta}
\def\bu{{\boldsymbol u}}
\def\urhat{\mathbf{\hat{r}}}
\newcommand{\br}{{\boldsymbol r}}
\newcommand{\bx}{{\boldsymbol x}}
\newcommand{\hr}{{\hat {\bf r}}}
\begin{document}
\title{Oscillations Modulating Power Law Exponents in Isotropic Turbulence:\\
Comparison of Experiments with Simulations}
\author {Kartik P. Iyer} 
\affiliation {Department of Physics, Department of Mechanical Engineering-Engineering Mechanics, Michigan Technological University, Houghton, MI $49931$}
\affiliation {Department of Mechanical and Aerospace Engineering, New York University, New York, NY, $11201$, USA}

\author {Gregory P. Bewley}
\affiliation {Sibley School of Mechanical and Aerospace Engineering, Cornell University, Ithaca, New York $14853$, USA}

\author {Luca Biferale}
\affiliation {Department of Physics and INFN, University of Rome Tor Vergata, Rome, $00133$, Italy}

\author {Katepalli R. Sreenivasan} 
\affiliation {Department of Mechanical and Aerospace Engineering, New York University, New York, NY, $11201$, USA}
\affiliation {Department of Physics and the Courant Institute of Mathematical Sciences, New York University, New York, NY $11201$, USA}
\email{krs3@nyu.edu}

\author {P. K. Yeung}
\affiliation {Schools of Aerospace and Mechanical Engineering, Georgia Institute of Technology, Atlanta, GA $30332$, USA}

\date{Postprint version of the manuscript published in Phys. Rev. Lett. {\bf{126}}, 254501 (2021)}

\begin{abstract}
Inertial-range features of turbulence are investigated using data from experimental measurements of grid turbulence and direct numerical simulations of isotropic turbulence simulated in a periodic box, both at the Taylor-scale Reynolds number $\rel \sim 1000$. In particular, oscillations modulating the power-law scaling in the inertial range are examined for structure functions up to sixth order moments. The oscillations in exponent ratios decrease with increasing sample size in simulations though, in experiments, they survive at a low value of $4$ parts in $1000$ even after massive averaging. The two data sets are consistent in their intermittent character but differ in small but observable respects. Neither the scaling exponents themselves nor all the viscous effects are consistently reproduced by existing models of intermittency.
\end{abstract}
\maketitle
\noindent
Turbulent fluctuations on scales intermediate between the small scale $\eta$ and the large scale $L$, the so-called inertial range, are thought to conform to power-laws \citep{MY75,Fri95,SA97,ishihara09}. In particular, one writes 
\beq
\label{stfn.eq}
S_m(r) \equiv \langle \left[ \delta_r u  \right]^m\rangle \sim r^{\zeta_m}
\;, \eta \ll r \ll L\;,
\eeq 
where $\delta_r u = [\bu(\br+\bx) - \bu(\bx)]\cdot \hr$ is the longitudinal velocity increment, $m$ is the moment-order, $\urhat$ is a unit vector along vector $\mbf{r}$, $r$ denotes magnitude of $\mbf{r}$ and $\la \cdot \ra$ denotes a suitable average. Although the theoretical basis for Eq.~\ref{stfn.eq} exists only for $m=3$ \citep{K41}, it is empirically regarded as viable for other orders as well \citep{KRS98}. The power-law scaling of Eq.~\ref{stfn.eq}, apart from offering the allure of ubiquity \citep{GF92,barbasi05,sims08,stumpf12}, often allows a simplification of subsequent analysis (for example, see \citep{eyink95}).
\begin{figure}[ht]
\includegraphics[width=0.5\textwidth,center]{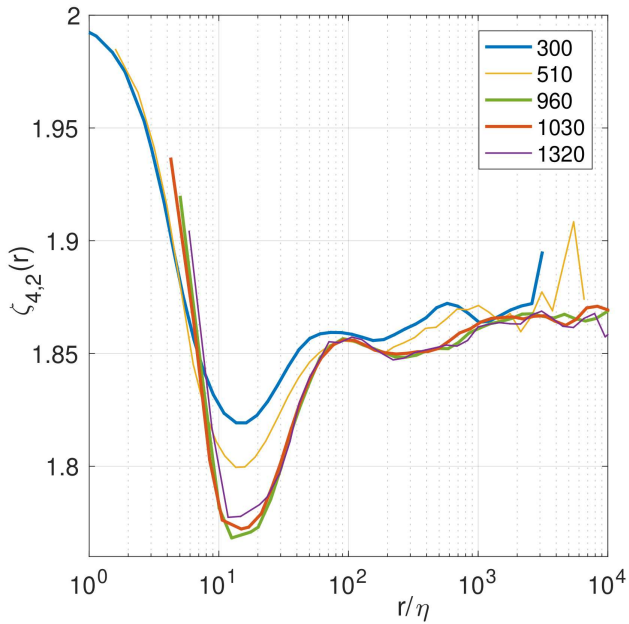}
\protect\caption{
Replot of data from~\citep{SBB17} of the ratio of logarithmic local slopes of fourth order longitudinal velocity structure function $(S_4)$ to the corresponding quantity for the second order $(S_2)$ {\it{vs}}.~spatial separation $r$ normalized by Kolmogorov scale $(\eta)$; see Eq.~(2). The two thinner curves were computed from datasets about ten times shorter than the others, as explained in \citep{SBB17}. The legend shows the microscale Reynolds numbers $\rel$.}
\label{sbb17.fig}
\end{figure}

\vspace{0.05cm}
Recently, extensively sampled data from grid turbulence \citep{SBB17} have shown an interesting feature with respect to Eq.~\ref{stfn.eq} (see Fig.~\ref{sbb17.fig}) that power law exponents may be modulated by undulations that are only partly explained by existing intermittency models \citep{Meneveau96}. These oscillations become explicit when the exponent ratios $\zeta_4/\zeta_2$ in Eq.~\ref{stfn.eq} are examined, as the authors of \citep{SBB17} showed (see Fig.~\ref{sbb17.fig}). The oscillations decrease with Reynolds number. Here, by examining even more extensive data from the same experiment \citep{SBB17} at one Reynolds number, along with those from direct numerical simulations \citep{Yeung15,KIKRS17,KIKRS20} at a comparable Reynolds number, we show that inertial range undulations observed in Fig.~\ref{sbb17.fig} manifest in exponent ratios such as $\zeta_4/\zeta_2$, in both experiments (EXP) and simulations (DNS), but diminish in the limit of massive averaging; they seem to disappear in simulations to yield pure power-law scaling (Eq.~\ref{stfn.eq}), but settle down to very small root-mean-square magnitudes (rms) of the order of $4$ parts in $1000$ in the experiment. The situation at higher $\rel$ is unknown at present. Both EXP and DNS show that the ratios $\zeta_4/\zeta_2$ and $\zeta_6/\zeta_2$ differ from the classical Kolmogorov phenomenology \citep{MY75}, but differ in small but observable respects between them. We comment on their possible origin. 
\vspace{0.05cm}
Table \ref{tab1} reports a few important parameters in EXP and DNS. The Reynolds numbers ($\rel$) in both EXP and DNS are sufficiently high and comparable to each other. The experiments were run for an unprecedented duration and averages performed over more than $10^5$ independent turn-over times of the turbulence, but the averaging is performed for only one component of the velocity and do not take account of possible residual anisotropies; the DNS data, on the other hand, extensive though they are, do not correspond to similarly large number of independent realizations but perform spherical averaging over the solid angle to obtain the isotropic sector \citep{KS00,BP05} by the method given in \citep{KI17}, and eliminates residual anisotropy effects.
\begin{figure}[ht]
\includegraphics[width=0.5\textwidth,center]{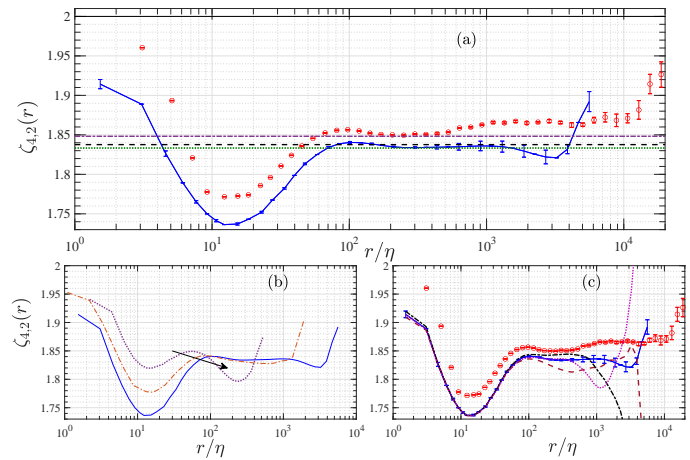}
\protect\caption{Ratio of logarithmic local slopes of $S_4$ and $S_2$, $\zeta_{4,2}(r) \equiv d[\log S_4(r)]/d[\log S_2(r)] $, versus spatial separation $r$ normalized by Kolmogorov scale $\eta$. (a) Data from EXP at $\rel=1030$ (open circles) are from \citep{SBB17}; the isotropic sector from DNS at $\rel=1300$ (solid line) \citep{KIKRS20} are compared with the $p$-model (dash-dot line) of Meneveau and Sreenivasan \citep{CMKRS87}, She-Leveque model (dashed line) \citep{SL94} and that by Yakhot (dotted line) \citep{Yakhot01}. Horizontal line at $\zeta_{4,2}=2$ corresponds to non-intermittent scaling \citep{K41}. Error bars indicate the standard error obtained from temporal fluctuations of local slopes. (b) Isotropic DNS data at different Reynolds numbers: $\rel = 240$ (dotted line), $650$ (dash-dot line) and $1300$ (solid line) are plotted to show that the viscous bottleneck around $\nr=80$ decreases in amplitude with increasing $\rel$ in the direction shown. (c) Data from EXP (open circles), compared with the DNS data for the isotropic sector (solid line) and the one-dimensional cuts in DNS along the three Cartesian directions: $\urhat = (1,0,0)$ (dotted line), $\urhat = (0,1,0)$ (dashed line) and $\urhat = (0,0,1)$ (dash-dot line). The latter three illustrate that the inertial range value of $\zeta_{4,2}(r)$ can be affected by non-universal large scale effects when not projecting onto the isotropic sector.}
\label{ess4.fig}
\end{figure}
\vspace{0.05cm}
In order to measure exponents $\zeta_m$ in Eq.~\ref{stfn.eq} we consider the logarithmic local slopes of velocity difference moments $S_m(r)$ of order $m$, 
\beq
\zeta_m(r) \equiv \frac{d \, \log S_m(r)}{d \, \log (r)} \;.
\label{ls.eq}
\eeq
If the moments $S_m(r)$ exhibit proper inertial range scaling according to Eq.~\ref{stfn.eq} then $\zeta_m(r)$ are constants in the inertial range $\eta \ll r \ll L$. In Fig.~\ref{ess4.fig} we compare the ratio of local slopes $\zeta_{4,2} \equiv d[\log S_4]/d[\log S_2] = \zeta_4(r)/\zeta_2(r)$ from EXP and DNS. This is the first non-trivial ratio between exponents of analytic functions that involve no modulus, and is less affected by strong cancellations (as can happen for odd-orders) and poor statistics (as can happen for higher-order moments) \citep{benzi93,benzi10}. The point of \citep{SBB17} was this ratio had an undulating character in the inertial range, albeit of decreasing magnitude with increasing Reynolds numbers. 

\begin{table}
    \begin{tabular}{|c|c|c|c|c|c|c}
     & $N^3$ & $L/\eta$ &   $ns$& $\rel$ \\
     \hline
    EXP & $\ldots$ & $2567$ &  $ 10^{10}$ & $1030$ \\
    DNS & $8192^3$ &$2514$& $10^{15}$ & $1300$ \\
    \end{tabular}
    \caption{Parameters of the data from the experiment and direct numerical simulation. $N^3$ is the number of collocation points for DNS in a periodic cube of fixed size $L_0$. $L/\eta$ is the ratio of the integral scale $L$ ($L/L_0 \approx 0.2$ for DNS) to the Kolmogorov scale $\eta \equiv (\nu^3/\epsm)^{1/4}$, where $\nu$ is the kinematic viscosity, $\epsm$ the mean dissipation rate; $ns$ is the total number of samples in space and/or time (but the meaning of $ns$ in EXP and DNS is different because the ratio of the number of independent samples to $ns$ differ between EXP and DNS); $\rel$ is the Taylor microscale Reynolds number.}
    \label{tab1}
\end{table}

Figure \ref{ess4.fig}(a) shows that the general trend in EXP (circles) and the isotropic DNS data (solid line) is very similar with a conspicuous $\rel$-dependent viscous dip around $r/\eta \sim 10$ predicted by the multifractal model  in both Eulerian and Lagrangian frameworks \citep{SBB17,FV91,Meneveau96,arneodo2008}. With increasing scale-size a crossover ``bottleneck" between the viscous and inertial regimes around $r/\eta = 80$ is seen in both EXP and DNS. This is not predicted by any existing models. This viscous bottleneck (which lies outside the inertial range) decreases in amplitude with increasing $\rel$ as shown in Fig.~\ref{ess4.fig}(b) in the DNS (see also Fig.~\ref{sbb17.fig}). The relation between this physical space bottleneck and that in the Fourier space \citep{DS10,kuchler19} remains to be understood properly. 

It is evident from Fig.~\ref{ess4.fig}(a) that inside the inertial range $\eta \ll r \ll L$, which is roughly estimated to be in the range $r/\eta \in (100,1000)$ \citep{KIKRS20}, $\zeta_{4,2}$ from both EXP and isotropic sector of DNS differ from the self-similar value of $ \zeta_{4,2}=2$ indicating that higher-$\rel$ turbulence is indeed intermittent with a definite departure from the K41-similarity \citep{K41}. One can also see that EXP and isotropic DNS data differ from one another and both show some $r$-dependent undulations instead of the $r$-independent constant (see Eq.~\ref{stfn.eq}). The finite (and small) mismatch in $\zeta_{4,2}$ between EXP and the isotropic sector from DNS could arise from the fact that the former is not projected onto the isotropic sector. Indeed, similar one-dimensional cuts in DNS shown in Fig.~\ref{ess4.fig}(c) behave differently along different directions beyond $\nr = 100$ with that along $\urhat =(0,0,1)$ matching the EXP data closely in the inertial range. Another possible reason for the mismatch might be that EXP relies on Taylor's frozen flow hypothesis \citep{taylor} while the DNS data do not. Finally, one cannot exclude non-universal and Reynolds-number-independent effects induced by different forcing mechanisms \citep{yakhot2018}.

We compare $\zeta_{4,2}$ in Fig.~\ref{ess4.fig}(a) with three different phenomenological models. While the $p$-model of Meneveau and Sreenivasan with the parameter $p$ set to their value of $0.7$ \citep{CMKRS87} compares favorably with EXP, the model by Yakhot \citep{Yakhot01,JS2007} closely matches DNS. The She-Leveque prediction \citep{SL94} lies in between EXP and DNS, being closer to the latter than the former. A similar comparison of $\zeta_{4,2}$ from EXP and DNS with other inertial range models \citep{benzi84,Ruelle12,oz15}, although not shown here, reveals that the agreement is qualitatively similar to those shown in Fig.~\ref{ess4.fig}.
\begin{figure}
\includegraphics [width=0.5\textwidth,center]{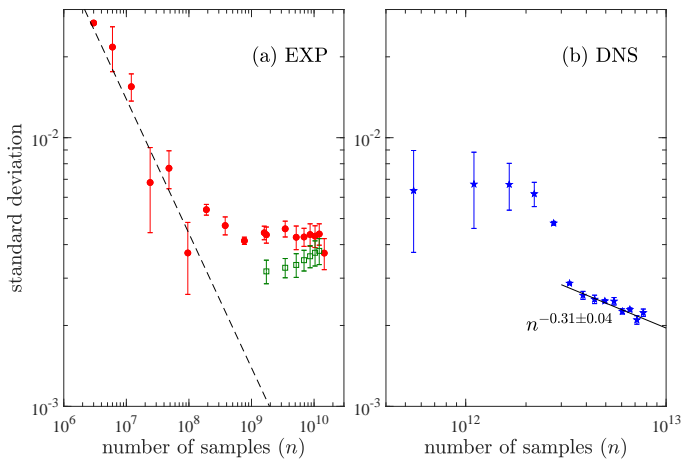}
\protect\caption{Log-log plot of the standard deviation of the oscillation amplitude in $d[\log(S_4)]/d[\log(S_2)]$ from its mean value in the range $r/\eta \in (100,1000)$ in Fig.~\ref{ess4.fig} as function of sample size $n$ for (a) the experiments and (b) the isotropic sector from DNS. Error bars indicate the variation obtained by changing the interval $r/\eta \in [100,1000]$ by $10\%$ on either side. We include an analysis of a second EXP dataset (green squares) acquired at $\rel=961$ in order to illustrate typical variation in large $n$ behavior. The standard deviation in the experiment is approximately constant for about two orders of magnitude for $n > 10^8$, or equivalently for more than $10^5$ turnover times, while that in DNS shows a $-0.31 \pm 0.04$ power-law decay for $n > 3 \times 10^{12}$, as indicated by the solid line. The sample size in DNS cannot be easily translated to turnover time scales. The dashed line in panel (a) corresponds to the $-1/2$ scaling of random noise.}
\label{conv.fig}
\end{figure}

We now assess in Fig.~\ref{ess4.fig} the $r$-dependent inertial range oscillations modulating the power-law expectations of Eq.~\ref{stfn.eq}, which is the feature to which Ref.~\citep{SBB17} drew attention. Noting that the power-law behavior of Eq.~\ref{stfn.eq} is expected to truly hold only at sufficiently large $\rel$ and as sample size $n \to \infty$, and that real systems such as the ones examined here are obviously at finite Reynolds numbers, we measure the oscillation amplitude for EXP and DNS and plot their standard deviations as functions of their respective sample size $n$ in Fig.~\ref{conv.fig}. EXP and DNS show different behaviors, with EXP suggesting a saturation at a small but finite value at even the large sample sizes considered here, while DNS exhibit a power-law decay with the sample size, even though care must be taken because of the short scaling range. Furthermore, the departure from a pure power law can have different origins depending on whether we are close to the viscous scale or to the integral scale. For example, while viscous effects are much less affected by statistical sampling, large-scale properties can be less stable because of lack of statistics in DNS and some sustained low-level forcing of a different kind in EXP. The differences in the data might also result from differences in the averaging procedures in the two instances or to differences in the extent of the two datasets. We recall that the standard deviation for random noise varies as $n^{-1/2}$ with sample size.
\begin{figure}
\includegraphics[width=0.5\textwidth,center]{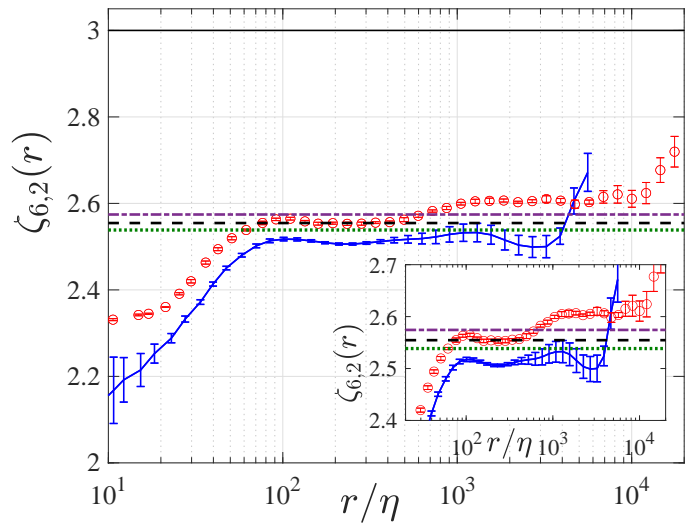}
\protect\caption{Ratio of logarithmic local slopes of $S_6$ and $S_2$, $\zeta_{6,2}(r)\equiv d[\log S_6(r)]/d[\log S_2(r)]$ versus spatial separation $r$ normalized by Kolmogorov scale $\eta$. Error bars indicate standard error obtained from temporal fluctuations of local slopes. Data from experiment (open circles) and the isotropic sector of DNS (solid line) are compared with the $p$-model (dash-dot line) of Meneveau and Sreenivasan \citep{CMKRS87}, She-Leveque model (dashed line) \citep{SL94} and that by Yakhot (dotted line) \citep{Yakhot01}. Horizontal line at $\zeta_{6,2} = 3$ corresponds to self-similar non-intermittent scaling \citep{K41}. Inset shows a blow-up of the inertial range and to either side of it to highlight differences between EXP and DNS.
}
\label{ess6.fig}
\end{figure}

\vspace{0.05cm} 
The differences that exist between EXP and DNS in the inertial range in Fig.~\ref{ess4.fig} persist at higher orders. Figure~\ref{ess6.fig} shows $\zeta_{6,2} \equiv d[\log S_6]/d[\log S_2] =\zeta_6/\zeta_2$ as a function of scale $r$ with a focus in the inertial range, $r/\eta \in (100,1000)$. K41-similarity \citep{K41} would correspond to $\zeta_{6,2} = 3$. Both EXP and isotropic DNS show roughly double the deviation from K41-similarity at the level of $\zeta_{6,2}$ than that at $\zeta_{4,2}$ (the increased departure being a characteristic of intermittency), and exhibit qualitatively similar behavior including the persistence of differences between EXP and DNS (see inset of Fig.~\ref{ess6.fig}). In the case of $\zeta_{6,2}$ the She-Leveque model \citep{SL94} seems to agree somewhat better with EXP than the $p$-model, which under predicts the level of intermittency. All three models appear to under-predict the inertial range intermittency of the DNS for $\zeta_{6,2}$.
  
\vspace{0.05cm}
In summary, we have examined the commonly held belief that non-universal large-scale and viscous effects are forgotten to yield universal statistics well inside the inertial range. In particular, we compared turbulence data from simulations and experiments at similar Reynolds numbers and examined if power-law scaling prevails in the inertial range and, if it does, to what extent the assumed universality \citep{SAW2018,peinke18} holds for structure functions. We confirmed  that the scaling exponent ratios depart considerably from the K41 prediction \citep{MY75,Fri95} and that existing phenomenological models do not account for all the observed small-scale non-universalities. We found that viscous effects in the exponent ratios persist up to at least  $r/\eta \sim 80$ (although this precise number may vary with the context). For larger scale separations and up to $r/\eta \sim 1000$ scaling properties are close to a pure power-law with superposed small oscillations whose amplitudes decrease with increasing scale and Reynolds number; they decay continually with sample size in DNS but have a sustained presence of about $4$ parts in $1000$ even for very large sample sizes in EXP. A small but detectable mismatch between DNS and EXP data is further measured at all scales. Both these effects may be due to Reynolds-number-independent breaking of universality, which would call to question the last $80$ years of turbulence theory, or to some subtle and long-lived large-scale effects, or something else. These important findings, though subtle, have been possible because of the unprecedented combination of high sample sizes and high Reynolds numbers in both DNS and EXP.  We are entering an era where scaling laws in turbulent flows can be assessed to within a few percent accuracy and over two or more decades of scale-by-scale comparisons.

\section{Acknowledgments}
The experiments were performed at the Max Planck Institute with M. Sinhuber and E. Bodenschatz, to whom we are grateful for the possibility to use the data. We thank D. Buaria for his contributions to the simulations.
This work is partially supported by the National Science Foundation
(NSF), via Grant No. ACI-$1640771$
at the Georgia Institute of Technology. This work used the Extreme Science and Engineering Discovery Environment (XSEDE), which is supported by National Science Foundation grant number ACI-1548562, on Stampede2 provided through Allocation TG-MCA99S022. The computations were performed using supercomputing resources through the Blue Waters Project at the National Center for Supercomputing Applications at the University of Illinois (Urbana-Champaign).
This work received funding from the European Research Council (ERC) under the European Union's Horizon 2020 research and innovation program (grant agreement No. 882340).


\end{document}